\documentclass[%
 reprint,
 amsmath,amssymb,
 aps,
]{revtex4-1}

\usepackage{graphicx}
\usepackage{dcolumn}
\usepackage{bm}
\usepackage{dsfont}
\usepackage{appendix}

\usepackage{color}
\usepackage{ulem}

\begin{document}

\preprint{APS/123-QED}

\title{Anomaly of a gauge theory under rescaling of the fields}

\author{Renata Jora
	$^{\it \bf a}$~\footnote[1]{Email:
		rjora@theory.nipne.ro}}
\email[ ]{rjora@theory.nipne.ro}

\affiliation{$^{\bf \it a}$ National Institute of Physics and Nuclear Engineering PO Box MG-6, Bucharest-Magurele, Romania}

\begin{abstract}

We determine the anomaly associated to an arbitrary scaling of the fields in a quantum gauge theory without making use of the Fujikawa method. We show that this anomaly is dependent on the spin term present in the  action and at one loop can be directly extracted from the spin contribution to the one loop effective action. Our results can be readily applied to any gauge theory, supersymmetric or not and agree with previous determination for supersymmetric gauge theories based on the Fujikawa method.

\end{abstract}
\maketitle

Quantum gauge field theories have an anomaly associated to the scale transformation given by:
\begin{eqnarray}
\theta^{\mu}_{\mu}=\frac{\beta(g)}{2g}F^{a\mu\nu}F^a_{\mu\nu},
\label{res43552}
\end{eqnarray}
where $F^a_{\mu\nu}$ is an abelian gauge tensor, $\beta(g)$ is the beta function of the coupling constant g and $\theta^{\mu}_{\mu}$ is the trace of the symmetric energy momentum tensor $\theta^{\mu}_{\nu}$.

The scale anomaly  \cite{Hata}-\cite{Tarasov} (and references therein) measures the behavior of the gauge theory under the scale transformation  which includes the scaling of space-time and also specific transformations for each field.  It might be useful to determine also the anomalous contribution to theory given by only the arbitrary scaling of the fields. Such a scaling was used earlier for supersymmetric gauge theories in \cite{Murayama} to connect the holomorphic and the canonical coupling constants. It turned out that such an endeavour was by no means trivial.  Here we will use a simple tractable method that makes no reference to the Fujikawa method employed usually. We will show  that although our derivation refers to QCD  our results may be directly applied to any gauge theory supersymmetric or not for which the one loop contribution (or higher orders) to the effective action is known.

Consider for illustration QCD based on the group $SU(N)$ and with $N_f$ Dirac fermions in the fundamental representation of the gauge group. The partition function corresponding to it is:
\begin{eqnarray}
Z=\int d A^a_{\mu}d \Psi^{ib} d\bar{\Psi}_{jc}dc^e d\bar{c}^f \exp[-i\int d^4 x {\cal L}].
\label{partfunc6453}
\end{eqnarray}
Here $i,j$ are flavors indices and all the other are color indices corresponding to the representation for each field. As usually $c$ and $\bar{c}$ denote the ghosts.

We make the following change of variables of integration in the partition function:
\begin{eqnarray}
&&A^a_{\mu}(x) \rightarrow A^{a\prime}_{\mu}(x')=A^a_{\mu}(x)-\alpha[A^a_{\mu}(x)+x^{\rho}\partial_{\rho}A_{\mu}(x)]
\nonumber\\
&&\Psi(x)\rightarrow \Psi'(x')=\Psi(x)-\alpha[\frac{3}{2}\Psi(x)+x^{\rho}\partial_{\rho}\Psi(x)]
\nonumber\\
&&\bar{\Psi}\rightarrow \bar{\Psi}^{\prime}(x')=\bar{\Psi}(x)-\alpha[\frac{3}{2}\bar{\Psi}(x)+x^{\rho}\partial_{\rho}\bar{\Psi}(x)]
\label{transf4553666}
\end{eqnarray}
One can recognize in the transformation in Eq. (\ref{transf4553666}) the standard scale transformation of the fields. A similar transformation may be associated to the ghosts or not.

Eq. (\ref{transf4553666}) introduces an jacobian in the partition function in Eq. (\ref{partfunc6453}):

\begin{eqnarray}
&&J(A^a_{\mu})=\det \Bigg[\frac{\delta A^a_{\mu}(x)}{\delta A^{b\prime}_{\rho}(x')}\Bigg]
\nonumber\\
&&J^{-1}(\Psi_{ib})=\det\Bigg[\frac{\delta \Psi_{ib}(x)}{\delta \Psi^{\prime}_{jc}(x')}\Bigg]
\nonumber\\
&&J^{-1}(\bar{\Psi}_{ib})=\det\Bigg[\frac{\delta \bar{\Psi}_{ib}(x)}{\delta \bar{\Psi}^{\prime}_{jc}(x')}\Bigg].
\label{jacob647888}
\end{eqnarray}
Here we took into account the anti-commuting nature of the fermion fields.

Next we need to consider how the action transforms. For that we write:
\begin{eqnarray}
&&\int d^4 x{\cal L}(A^a_{\mu}(x),\Psi_{ib}(x),\bar{\Psi}_{jc}(x))=
\nonumber\\
&&\int d^4 x' {\cal L}((A^a_{\mu}(x'),\Psi_{ib}(x'),\bar{\Psi}_{jc}(x'))=
\nonumber\\
&&\int d^4 x' {\cal L}((A^{\prime a}_{\mu}(x'),\Psi'_{ib}(x'),\bar{\Psi}'_{jc}(x'))+
\nonumber\\
&&\int d^4 x'\Bigg[{\cal L}((A^a_{\mu}(x'),\Psi_{ib}(x'),\bar{\Psi}_{jc}(x'))-
\nonumber\\
&& {\cal L}((A^{\prime a}_{\mu}(x'),\Psi'_{ib}(x'),\bar{\Psi}'_{jc}(x'))\Bigg].
\label{lagrch65777}
\end{eqnarray}

Then the change in the action given by the big square bracket in Eq. (\ref{lagrch65777}) corresponds to the contribution from the jacobians in Eq. (\ref{jacob647888}). One can easily determine the quantity in the square bracket as the integral in space time of:
\begin{eqnarray}
\alpha[T^{\mu}_{\mu}-\theta^{\mu}_{\mu}],
\label{sqresult65775}
\end{eqnarray}
where $\theta^{\mu}_{\mu}$ is the trace of the symmetric energy tensor and $T^{\mu}_{\mu}$ is the trace of the canonical energy momentum tensor according to:
\begin{eqnarray}
&&\theta^{\mu}_{\mu}=
\nonumber\\
&&\partial_{\mu}\Bigg[\frac{\partial {\cal L}}{\partial \partial_{\mu}\Phi_i}\delta(\Phi_i)\Bigg]+
\Bigg[\frac{\partial {\cal L}}{\partial \partial_{\mu}\Phi_i}\partial_{\mu}\Phi_i-4{\cal L}\Bigg]=
\nonumber\\
&&\partial_{\mu}\Bigg[\frac{\partial {\cal L}}{\partial \partial_{\mu}\Phi_i}\delta(\Phi_i)\Bigg]+ T^{\mu}_{\mu}.
\label{tensors65774}
\end{eqnarray}
Here $\Phi_i$ denote any generical field present in the action, gauge or fermion ones. Then a change due to only scaling of the fields is given by:
\begin{eqnarray}
\alpha[T^{\mu}_{\mu}-\theta^{\mu}_{\mu}]=\alpha\partial_{\mu}\Bigg[\frac{\partial {\cal L}}{\partial \partial_{\mu}\Phi_i}\delta(\Phi_i)\Bigg].
\label{resinv6564}
\end{eqnarray}
The expression on the right hand side of Eq. (\ref{resinv6564}) looks complicated and unhelpful. However in \cite{trace} a more amenable version for a general gauge theory was introduced in the form:
\begin{eqnarray}
T^{\mu\nu}-\theta^{\mu\nu}=-\partial_{\rho}\chi^{\mu\rho\nu},
\label{epr76886}
\end{eqnarray}
where:
\begin{eqnarray}
\chi^{\mu\rho\nu}=-2\frac{\partial {\cal L}_g}{\partial F^a_{\mu\rho}}A^{a\nu}.
\label{derivexpr65775}
\end{eqnarray}
Here ${\cal L}_g$ is the gauge invariant kinetic term for the gauge fields.

Next we will determine the exact contribution of the tensor in Eq. (\ref{derivexpr65775}). We will work in the background gauge field method where the gauge field become $A^a_{\mu}\rightarrow B^a_{\mu}+A^a_{\mu}$ where $B^a_{\mu}$ is a background gauge field and $A^a_{\mu}$ becomes the fluctuating field. The gauge kinetic part of the Lagrangian  in the background gauge field is \cite{Peskin}:
\begin{eqnarray}
-\frac{1}{4g^2}\Bigg[B^a_{\mu\nu}+D_{\mu}A^a_{\nu}-D_{\nu}A^a_{\mu}+f^{abc}A^b_{\mu}A^c_{\nu}\Bigg]^2,
\label{kinterms64553}
\end{eqnarray}
where $B^a_{\mu\nu}$ is the gauge tensor for the background gauge field. Since we do not scale the background gauge field we need to consider the tensor $\chi^{\mu\rho\nu}$ pertaining only to the fluctuating field. Then it may be written as:
\begin{eqnarray}
\chi_{\mu\rho}^{\mu}=-2\frac{\partial {\cal L}_g}{\partial F^{a\mu\rho}}[A^{a\mu}+B^{a\mu}]+2\frac{\partial {\cal L}}{\partial B^{a\mu\rho}}[B^{a\mu}].
\label{expr76885999}
\end{eqnarray}
Here we consider the tensor for the full Lagrangian and extracted the contribution from the background gauge field. Eq. (\ref{expr76885999}) can be written further as:
\begin{eqnarray}
&&\chi_{\mu\rho}^{\mu}=
\nonumber\\
&&\frac{1}{g^2}F^a_{\mu\rho}[A^{a\mu}+B^{a\mu}]-\frac{1}{g^2}F^a_{\mu\rho}B^{a\mu}+2\frac{\partial {\cal L}_m}{\partial B^{a\mu\rho}}[B^{a\mu}]=
\nonumber\\
&&\frac{1}{g^2}F^a_{\mu\rho}A^{a\mu}+2\frac{\partial {\cal L}_m}{\partial B^{a\mu\rho}}[B^{a\mu}]=
\nonumber\\
&&\frac{1}{g^2}B^a_{\mu\rho}A^{a\mu}+\frac{1}{g^2}[D_{\mu}A^a_{\rho}-D_{\rho}A^a_{\mu}]A^{a\mu}+
\nonumber\\
&&\frac{1}{g^2}f^{abc}A^b_{\mu}A^c_{\rho}A^{a\mu}+2\frac{\partial {\cal L}_m}{\partial B^{a\mu\rho}}[B^{a\mu}].
\label{calc75664}
\end{eqnarray}
In the background gauge field formalism the contribution from the gauge and fermion terms at one loop to the effective action are stemming from\cite{Peskin}:
\begin{eqnarray}
&&{\cal L}_A=-\frac{1}{2g^2}\Bigg[A^a_{\mu}[-(D^2)^{ac}g^{\mu\nu}-2f^{abc}B^{b\mu\nu}]A^c_{\nu}\Bigg]
\nonumber\\
&&\Bigg[\det[i\gamma^{\mu}D_{\mu}]^2\Bigg]^{\frac{N_f}{2}}=\Bigg[\det[-D^2+B^{b}_{\rho\sigma}S^{\rho\sigma}t^b]\Bigg]^{\frac{N_f}{2}},
\label{contr65776}
\end{eqnarray}
where in the last line we considered the integral of the fermion quadratic term. One notices immediately that on the right hand side of Eq. (\ref{calc75664}) the first term must be dropped as being linear in the fluctuating field and the third term must also be dropped because it is trilinear so it will bring contribution only at two loops. We are interested only in the one loop result as the full contribution might contain terms at any loop order (as opposed to the supersymmetric QCD where we expect that the full contribution is only at one loop). With regard to the contribution of the gauge fluctuating fields we are left only with the second term and the last term in the last line of Eq. (\ref{calc75664}). The second term may be written as:
\begin{eqnarray}
&&\frac{1}{g^2}[D_{\mu}A^a_{\rho}-D_{\rho}A^a_{\mu}]A^{a\mu}=
\nonumber\\
&&\frac{1}{g^2}\Bigg[\partial_{\mu}A^a_{\rho}+f^{abc}B^b_{\mu}A^c_{\rho}-\partial_{\rho}A^a_{\mu}-f^{abc}B^b_{\rho}A^a_{\mu}\Bigg]A^{a\mu}=
\nonumber\\
&&\frac{1}{g^2}\Bigg[\partial_{\mu}A^a_{\rho}-\partial_{\rho}A^a_{\mu}\Bigg]A^{a\mu}+
\nonumber\\
&&\frac{1}{g^2}\Bigg[f^{abc}B^b_{\mu}A^c_{\rho}-f^{abc}B^b_{\rho}A^a_{\mu}\Bigg]A^{a\mu}.
\label{finres648899}
\end{eqnarray}

We introduce the result in the last line of Eq. (\ref{finres648899}) into  Eq. (\ref{epr76886}) to obtain;
\begin{eqnarray}
&&[T^{\mu}_{\mu}-\theta^{\mu}_{\mu}]_A=-[\partial^{\rho}\chi_{\mu\rho}^{\mu}]_A=
\nonumber\\
&&-\partial^{\rho}\frac{1}{g^2}\Bigg[\partial_{\mu}A^a_{\rho}-\partial_{\rho}A^a_{\mu}\Bigg]A^{a\mu}-
\nonumber\\
&&-\partial^{\rho}\frac{1}{g^2}\Bigg[f^{abc}B^b_{\mu}A^c_{\rho}-f^{abc}B^b_{\rho}A^c_{\mu}\Bigg]A^{a\mu}.
\label{reexpect65774}
\end{eqnarray}
The subscript $A$ refers to the fluctuating gauge fields contribution. The expression in the second line is a total derivative which does not contain the background gauge field so in the quantum approach will lead to zero. The second term on the last line is zero by antisymmetry. The first term on the last line  may be calculated to lead to:
 \begin{eqnarray}
&&\alpha(x)\Bigg[T^{\mu}_{\mu}-\theta^{\mu}_{\mu}\Bigg]_A=-\alpha(x)[\partial^{\rho}\chi_{\mu\rho}^{\mu}]_A=
\nonumber\\
&&[\partial^{\rho}(\alpha(x)]\frac{1}{g^2}\Bigg[f^{abc}B^b_{\mu}A^c_{\rho}A^{a\mu}\Bigg].
\label{res8677465}
\end{eqnarray}
Next we need to compute the contribution of the last term in the background gauge field method at one loop. Before doing that we need to calculate the similar contribution from the fermion fields using  Eq. (\ref{calc75664}) and the second line in (\ref{contr65776}):
\begin{eqnarray}
&&\frac{\partial {\cal L}_m}{\partial B^{a\mu\rho}}[B^{a\mu}]=
\nonumber\\
&&\frac{\partial}{\partial B^{a\mu\rho}}\ln\Bigg[\det[-D^2+B^{b\rho\sigma}S_{\rho\sigma}t^b]\Bigg]^{\frac{1}{2}}B^{a\mu}\approx
\nonumber\\
&&\frac{1}{2}(-\partial^2)^{-1}B^{a\mu}S_{\mu\rho}t^b.
\label{contr65774}
\end{eqnarray}
Note that contribution of the fermions is written schematically in terms of operators and we consider terms that might contribute only at one loop. Moreover the estimate is made for one single flavor of fermions.  Then:
\begin{eqnarray}
&&\alpha(x)\Bigg[T^{\mu}_{\mu}-\theta^{\mu}_{\mu}\Bigg]_f=-\alpha(x)[\partial^{\rho}\chi_{\mu\rho}^{\mu}]_f=
\nonumber\\
&&[\partial^{\rho}\alpha(x)](-\partial^2)^{-1}B^{a\mu}S_{\mu\rho}t^b,
\label{contr65774}
\end{eqnarray}
again written in terms of operators.

 We observe that in both equations (\ref{res8677465}) and (\ref{contr65774}) appear spin contribution which at one loop may be associated only with the spin term in Eq. (\ref{contr65776}). There is  no need to compute explicitly any of the integrals. In the background gauge field method one knows that the contribution of the spin operators comes from:
\begin{eqnarray}
-\frac{1}{2}{\rm Tr}\Bigg[(-\partial^2)^{-1}\Delta_J(-\partial^2)^{-1}\Delta_J],
\label{spi867758}
\end{eqnarray}
where,
\begin{eqnarray}
\Delta_J=B^b_{\rho\sigma}{\cal J}^{\rho\sigma}t^b,
\label{rez74665}
\end{eqnarray}
where ${\cal J}^{\rho\sigma}$ is the spin operator particular for each spin representation. Contribution of these spin terms to the scale anomaly (with the parameter $\alpha(x)$ of the effective Lagrangian in the background gauge field method) can be calculated as follows \cite{Peskin}:
\begin{eqnarray}
&&-\frac{1}{2}{\rm Tr}\Bigg[(-\partial^2)^{-1}\Delta_J(-\partial^2)^{-1}\Delta_J]=
\nonumber\\
&&-\frac{1}{2}{\rm Tr}\Bigg[(-\partial^2)_x^{-1}\delta(x-y)\Delta_J(y)(-\partial^2)_y^{-1}\delta(y-x)\Delta_J(x)]=
\nonumber\\
&&-\frac{1}{2}{\rm Tr} \int d^4 x \int d^4 y\int \frac{d^4 p}{(2\pi)^4}\int\frac{d^4 q}{(2\pi)^4}\int \frac{d^4 r}{(2\pi)^4}\int\frac{d^4 k}{(2\pi)^4}\times
\nonumber\\
&&\Bigg[\frac{1}{p^2}\exp[ip(x-y)]\Delta_J(k)\exp[iky]\times
\nonumber\\
&&\frac{1}{q^2}\exp[iq(y-x)]\Delta_J(r)\exp[irx]\Bigg]=
\nonumber\\
&&-\frac{1}{2}{\rm Tr}\int \frac{d^4 p}{(2\pi)^4}\int\frac{d^4 q}{(2\pi)^4}\int \frac{d^4 r}{(2\pi)^4}\int\frac{d^4 k}{(2\pi)^4}\times
\nonumber\\
&&\delta(p-q+r)\delta(-p+k+q)\times
\nonumber\\
&&\Bigg[\frac{1}{p^2}\exp[ip(x-y)]\Delta_J(k)\exp[iky]\times
\nonumber\\
&&\frac{1}{q^2}\exp[iq(y-x)]\Delta_J(r)\exp[irx]\Bigg]=
\nonumber\\
&&-\frac{1}{2}{\rm Tr}\Bigg[\int \frac{d^4k}{(2\pi)^4}\int \frac{d^4 q}{(2\pi)^4}\frac{1}{q^2}\frac{1}{(q+k)^2}\times
\nonumber\\
&&F^a_{\rho\sigma}(k)F^{b}_{\alpha\beta}(-k)t^at^b{\cal J}^{\rho\sigma}{\cal J}^{\alpha\beta}\Bigg]=
\nonumber\\
&&i\frac{1}{4}\int \frac{d^4k}{(2\pi)^4}F^a_{\mu\nu}(k)F^{a\mu\nu}(-k)\frac{4C(r)C(j)}{4\pi^2}\Gamma[2-\frac{d}{2}].
\label{rezfin5664}
\end{eqnarray}
Here we worked in dimensional regularization scheme and $C(j)$ is the result of summation over the spin structure operators ($C(j)=2$ for the gauge fields and $C(j)=1$ for the Dirac fields):
\begin{eqnarray}
{\rm Tr}[{\cal J}^{\rho\sigma}{\cal J}^{\alpha\beta}]=(g^{\rho\alpha}g^{\sigma\beta}-g^{\rho\beta}g^{\sigma\alpha})C(j).
\label{spinstructure54}
\end{eqnarray}
 Moreover $\delta^{ab}C(r)={\rm Tr}[t^at^b]$ where $t^a$ are the generators of the adjoint representation for the gauge fields ($C(r)=N$) and of the fundamental representation for the fermion fields ($C(r)=\frac{1}{2}$). The contribution to the effective action must be multiplied by the power of the quadratic  operators in the partition function as in Eq. (\ref{contr65776}).
Finally the relevant result for the spin contribution in the effective action is:
\begin{eqnarray}
-\alpha(x)\Bigg[\frac{8N}{64\pi^2}-\frac{2N_f}{64\pi^2}\Bigg]B^a_{\mu\nu}B^{a\mu\nu},
\label{res53442}
\end{eqnarray}
where here the role of $\alpha(x)$ is played by $\ln (k)$.

Then one can infer straightforwardly the corresponding terms coming from Eq. (\ref{res8677465})  and (\ref{contr65774}):
\begin{eqnarray}
&&\alpha(x)\Bigg[T^{\mu}_{\mu}-\theta^{\mu}_{\mu}\Bigg]=
\nonumber\\
&&\partial^{\rho}\alpha(x)B^{a\mu}F^{a\mu\rho}2\Bigg[\frac{8N}{64\pi^2}-\frac{2N_f}{64\pi^2}\Bigg]=
\nonumber\\
&&-\alpha(x)B^{a\mu\rho}B^a_{\mu\rho}\Bigg[\frac{8N}{64\pi^2}-\frac{2N_f}{64\pi^2}\Bigg],
\label{res677888}
\end{eqnarray}
where the contribution of gauge fields and fermions can be distinguished easily. Here the factor of $2$ in front of the square bracket and the absence of the minus sign come from the fact that the expansion of the operators is done only in the first order.

We go back to  Eq. (\ref{transf4553666}) to find the exact scaling that corresponds to the result in Eq. (\ref{res677888}). One has:
\begin{eqnarray}
&&A^{a\prime}_{\mu}(x')A^{a\prime}_{\nu}(x')=
\nonumber\\
&&A^a_{\mu}(x)A^a_{\nu}(x)-\alpha[2-\partial_{\rho}x^{\rho}]A^a_{\mu}(x)A^a_{\nu}(x)=
\nonumber\\
&&A^a_{\mu}(x)A^a_{\nu}(x)+2\alpha A^a_{\mu}(x)A^a_{\nu}(x)
\nonumber\\
&&\bar{\Psi}'(x')\Psi'(x')=\bar{\Psi}(x)\Psi(x)+\alpha\bar{\Psi}(x)\Psi(x).
\label{scaletransfr65774}
\end{eqnarray}
Here we took into account the Fujikawa approach where the fields appear in pairs and the result is correct up to a total derivative. Consequently the emergent scaling is $-1$ for the gauge fields and $-\frac{1}{2}$ for the fermion fields. For the same scaling  for fermions as for the gauge fields we need the multiply the fermion term in Eq. (\ref{res677888}) by 2. Moreover since the integration variables are in terms of prime fields we need to multiply the full result by (-1) which would correspond  to a natural scaling of the fields by $\alpha$. Then  the result of scaling by $\alpha$ of each field the Lagrangian is:
\begin{eqnarray}
&&\alpha B^{a\mu\rho}B^a_{\mu\rho}\Bigg[\frac{8N}{64\pi^2}-\frac{4N_f}{64\pi^2}\Bigg]=
\nonumber\\
&&\alpha \frac{1}{4}B^{a\mu\rho}B^a_{\mu\rho}\Bigg[\frac{4N}{8\pi^2}-\frac{2N_f}{8\pi^2}\Bigg].
\label{finalresult6577488}
\end{eqnarray}

In \cite{Murayama} the authors computed the contribution to the supersymmetric QCD Lagrangian coming from the scaling of the gauge fields by $g_c$ and from the scaling of the matter fields by $Z_f$ where $Z_f$ is the renormalization constant associated to the matter fields. Assuming $\ln Z_f=\ln g_c=\alpha$ the contribution reads:
\begin{eqnarray}
-\frac{1}{4}\alpha \Bigg[-\frac{N}{4\pi^2}+\frac{N_f}{4\pi^2}\Bigg]W^a_{\mu\nu}W^{a\mu\nu},
\label{supersym778}
\end{eqnarray}
where $W^a_{\mu\nu}$ is the supersymmetric gauge tensor that includes gauge fields and gluinos.

Let us write the result that would correspond to the supersymmetric Lagrangian in our approach. Because we deal with spin structures the scalar fields should not bring any contributions. Then in the result of Eq. (\ref{finalresult6577488}) the gluons and fermions would have exactly the same terms and we need to add only the gluino contribution. Since the gluinos are in the adjoint representations a factor of $2N$ should be added (which includes the absence of the factor $\frac{1}{2}$ and the group constant $N$). Moreover because the gluinos are Majorana fermions a factor of $\frac{1}{2}$ must be considered. Finally one obtains:
\begin{eqnarray}
&&\frac{1}{4}\alpha W^{a\mu\rho}W^a_{\mu\rho}\Bigg[\frac{4N}{8\pi^2}-\frac{2N}{8\pi^2}-\frac{2N_f}{8\pi^2}\Bigg]=
\nonumber\\
&&-\frac{1}{4}\alpha \Bigg[-\frac{N}{4\pi^2}+\frac{N_f}{4\pi^2}\Bigg]W^a_{\mu\nu}W^{a\mu\nu},
\label{coincide6}
\end{eqnarray}
result which coincides exactly to that in Eq. (\ref{supersym778}).

 One can apply the method introduced here to QCD in the background gauge field method to obtain a trivial results or to QCD in the regular renormalization method. However the latter requires and deserves a detailed treatment in a separate work due to the more complicated  relation between the renormalization  constants.

In order to show the relevance of our works we will consider  QED.
The renormalized QED Lagrangian is:
\begin{eqnarray}
&&{\cal L}=\sum_fZ_2\bar{\Psi}_fi\gamma^{\mu}\partial_{\mu}\Psi_f+Z_1e\sum_f\bar{\Psi}f\gamma^{\mu}A_{\mu}\Psi_f-
\nonumber\\
&&\frac{1}{4}Z_3F^{\mu\nu}F_{\mu\nu},
\label{QEDlagr23}
\end{eqnarray}
where $e_0Z_2Z_3^{1/2}=eZ_1$ and the bare index corresponds to the bare charge and the rest of the quantities are renormalized. Moreover for QED $Z_1=Z_2$. The sum is considered over fermion flavors $f$ which are all assumed with the same charge.

The beta function at two loops is:
\begin{eqnarray}
\beta(e)=\frac{N_f}{12\pi^2}e^3+\frac{N_f}{64\pi^4}e^5,
\label{betacga54}
\end{eqnarray}
whereas the renormalization constant $Z_2$ at one loop has the expression:
\begin{eqnarray}
Z_1=Z_2=1-\frac{e^2}{8\pi^2}\frac{1}{\epsilon}+...,
\label{const4655}
\end{eqnarray}
where $d=4-2\epsilon$ in dimensional renormalization scheme. We make the change of variables in the partition function $\Psi_f=Z_2^{-1/2}\Psi_f'$, $\bar{\Psi}_f=Z_3^{-1/2}\Psi_f'$ and $A_{\mu}=Z_3^{-1/2}A_{\mu}'$ (a initial scaling of the fields $e_0A_{\mu}=A_{\mu}$ is implicitly assumed). Then the effective action at one loop will be:
\begin{eqnarray}
-\frac{1}{4}\Bigg[\frac{1}{e_0^2}-\frac{N_f}{6\pi^2}\ln(k)-N_f\ln[Z_2]\frac{1}{4\pi^2}\Bigg]F^{\mu\nu}F_{\mu\nu}.
\label{partfeff45}
\end{eqnarray}
Here we applied the result in Eq. (\ref{finalresult6577488}) and took into account the fact that the sign is opposite. Then in order to establish the correct structure of the effective action one should have:
\begin{eqnarray}
\frac{1}{e_0^2}-\frac{N_f}{6\pi^2}\ln(k)-N_f\ln[Z_2]\frac{1}{4\pi^2}=\frac{1}{e^2}.
\label{resx5677}
\end{eqnarray}
We apply $\frac {d}{d\ln \frac{k}{M}}$ to Eq. (\ref{resx5677}) and use Eq. (\ref{const4655}) to obtain:
\begin{eqnarray}
-\frac{1}{6\pi^2}N_f-N_f\frac{e^2}{32\pi^2}=-2\frac{\beta(e)}{e^3},
\label{res66453663}
\end{eqnarray}
which evidently leads to the QED beta function at two loops as in Eq. (\ref{betacga54}).

Our calculations showed undoubtedly that for any gauge theory supersymmetric or not the anomaly associated to an arbitrary scaling of the fields can be readily associated and extracted at least at one loop from the spin dependent contribution of the fields to the one loop effective action.  The method employed here and the results may have important application in deciphering the properties of any gauge theory.


\begin{thebibliography}{30}

\bibitem{Hata} H. Hata, Progress of Theoretical Physics {\bf 65}, 1052-1057 (1981).
\bibitem{Fukikawa} K. Fujikawa, Phys. Rev. Lett. {\bf 44}, 1733 (1980).
\bibitem{Ferrara} S. Ferrara and B. Zumino, Nucl. Phys. B {\bf 87}, 207 (1975).
\bibitem{Clark} T. E. Clark, O. Piguet and K. Sibold, Nucl. Phys. B {\bf 143}, 445 (1978).
\bibitem{Piguet} O. Piguet and K. Sibold, Nucl. Phys. B {\bf 196}, 428 (1982); Nucl. Phys. B {\bf 196}, 447 (1982).
\bibitem{Tarasov} O. V. Tarasov and A. A. Vladimirov, Phys. Lett. B {\bf 96}, 94 (1980); M. T. Grisaru, M. Rocek and W. Siegel, Phys. Rev. Lett. {\bf 45}, 1063 (1980); W. Caswell and D. Zanon, Phys. Lett. B {\bf 100}, 152 (1980); L. V. Avdeev and D. Zanon, Phys. Lett. B {\bf 112}, 356 (1982).
\bibitem{Murayama}  N. Arkani-Hamed and H. Murayama, JHEP 006, 030 (2000); arXiv:hep-rh/9707133.
\bibitem{trace} D. N. Blaschke, F. Gieres, M. Reboud and M. Schweda, Nucl. Phys. B {\bf  912}, 192 (2016); arXiv:1605.01121.
\bibitem{Peskin} M. E. Peskin and D. V. Schroeder, "An Introduction to Quantum Field Theory", Westview Press Inc. (1995).












\end{thebibliography}
\end{document}